\documentclass[british]{elsarticle}
\usepackage[LGR,T1]{fontenc}
\usepackage[latin9]{inputenc}
\usepackage{verbatim}
\usepackage{url}
\usepackage{stackrel}
\usepackage{graphicx}

\makeatletter

\DeclareRobustCommand{\greektext}{%
  \fontencoding{LGR}\selectfont\def\encodingdefault{LGR}}
\DeclareRobustCommand{\textgreek}[1]{\leavevmode{\greektext #1}}
\ProvideTextCommand{\~}{LGR}[1]{\char126#1}

\providecommand{\tabularnewline}{\\}
\journal{Nucl.Instr.Meth. A}
\makeatother

\usepackage{babel}
\begin{document}

\begin{frontmatter}{}

\title{Simulation of the Response of the Solid State Neutron Detector for
the European Spallation Source}

\author[gla]{L. Boyd\fnref{mod}}

\author[lu]{E. Rofors}

\author[gla]{J. R. M. Annand\corref{cauth}}

\ead{john.annand@glasgow.ac.uk}

\author[lu,ess]{K. G. Fissum}

\author[ess,gla,milan]{R. Hall-Wilton}

\author[gla,ess]{R. Al Jebali}

\author[ess]{K. Kanaki}

\author[gla]{K. Livingston}

\author[lu,ess]{V. Maulerova}

\author[lu]{N. Mauritzson}

\author[gla]{R. Montgomery}

\author[lu,ess]{H. Perrey}

\author[gla]{B. Seitz}

\fntext[mod]{Present Address: HM Naval Base Clyde Off Site Centre, Rhu, Helensburgh, G84 8NE, UK.}


\address[gla]{School of Physics and Astronomy, University of Glasgow G12 8QQ, Scotland,
UK}

\address[lu]{Division of Nuclear Physics, Lund University, SE-221 00, Lund, Sweden}

\address[ess]{Detector Group, European Spallation Source ERIC, SE-221 00 Lund,
Sweden}

\address[milan]{Universita degli Studi di Milano-Bicocca, Piazza della Scienza 3,
20126, Milan, Italy}

\cortext[cauth]{Corresponding author}
\begin{abstract}
The characteristics of the Solid-state Neutron Detector, under development
for neutron-scattering measurements at the European Spallation Source,
have been simulated with a Geant4-based computer code. The code models
the interactions of thermal neutrons and ionising radiation in the
$^{6}$Li-doped scintillating glass of the detector, the production
of scintillation light and the transport of optical, scintillation
photons through the the scintillator, en route to the photocathode
of the attached multi-anode photomultiplier. Factors which affect
the optical-photon transport, such as surface finish, pixelation of
the glass sheet, provision of a front reflector and optical coupling
media are compared. Predictions of the detector response are compared
with measurements made with neutron and gamma-ray sources, a collimated
alpha source and finely collimated beams of 2.5 MeV protons and deuterons.
\end{abstract}

\end{frontmatter}{}

\section{Introduction}

The European Spallation Source (ESS) \citep{ess1,ess2} based in Lund,
Sweden is currently developing a number of instruments \citep{ess3}
for neutron science, one of which is the Small-K Advanced DIfractometer
(SKADI) \citep{skadi}. This small-angle neutron-scattering instrument
will be used to investigate the properties of materials used in a
broad range of scientific and medical research. The detector associated
with the SKADI instrument is the Solid-state Neutron Detector (SoNDe)
\citep{SoNDe,SoNDe2,SoNDe3} which has a total area of $\sim1040\times1040$~mm$^{2}$,
and is constructed from 400 $52\times52$~mm$^{2}$ elements. Each
element consists of a $50\times50\times1$~$\mathrm{mm^{3}}$ sheet
of GS20 \citep{GS20} $^{6}$Li-doped glass scintillator coupled optically
to a Hamamatsu H12700A $8\times8$ pixel, multi-anode photomultiplier
tube (MAPMT) \citep{MAPMT}. With this optical sensor, it is envisioned
that the neutron-interaction position resolution will be $\sim6$~mm,
the dimension of a single MAPMT pixel. The module is designed to detect
thermal neutrons through neutron capture within the scintillator $n\mathrm{+^{6}Li\rightarrow^{3}H+^{4}He}$.
The alpha particle and triton, with a total energy of 4.78~MeV, produce
optical scintillation photons which are detected by the MAPMT. A small
self-contained data-acquisition (DAQ) system \citep{ideas}, attached
directly to the rear of the MAPMT, digitises the MAPMT anode signals
to produce a data set containing the amplitude, position and timing
of the scintillation signal. 

Various configurations of the module including: 
\begin{itemize}
\item the use of a pixelated GS20 sheet 
\item the use of optical coupling medium
\item the provision of a front reflector
\item imperfect surface finish of the GS20 sheet
\end{itemize}
have been simulated to gauge their effect on the transport of optical
photons. The response of the detector module has been measured using
collimated beams of protons and deuterons \citep{Rofors_proton},
alpha particles \citep{Rofors-Alpha}, thermal neutrons \citep{n_IFE}
and uncollimated fast-neutron and gamma-ray sources. These measurements
are compared with the predictions of the computer model described
in this paper. 

\section{\label{sec:The-Computer-Model}The Computer Model}

The computer model of a SoNDe module uses the Geant4 Monte Carlo toolkit,
version 4.10.6, \citep{G4} and is coded in C++. Included in the model
are the GS20 sheet, the glass of the MAPMT window, the photocathode,
optional optical coupling media between the scintillator and MAPMT
and optional front reflectors.

\subsection{\label{subsec:Module-Geometry}Module Geometry}

\begin{figure}[h]
\includegraphics[width=1\columnwidth]{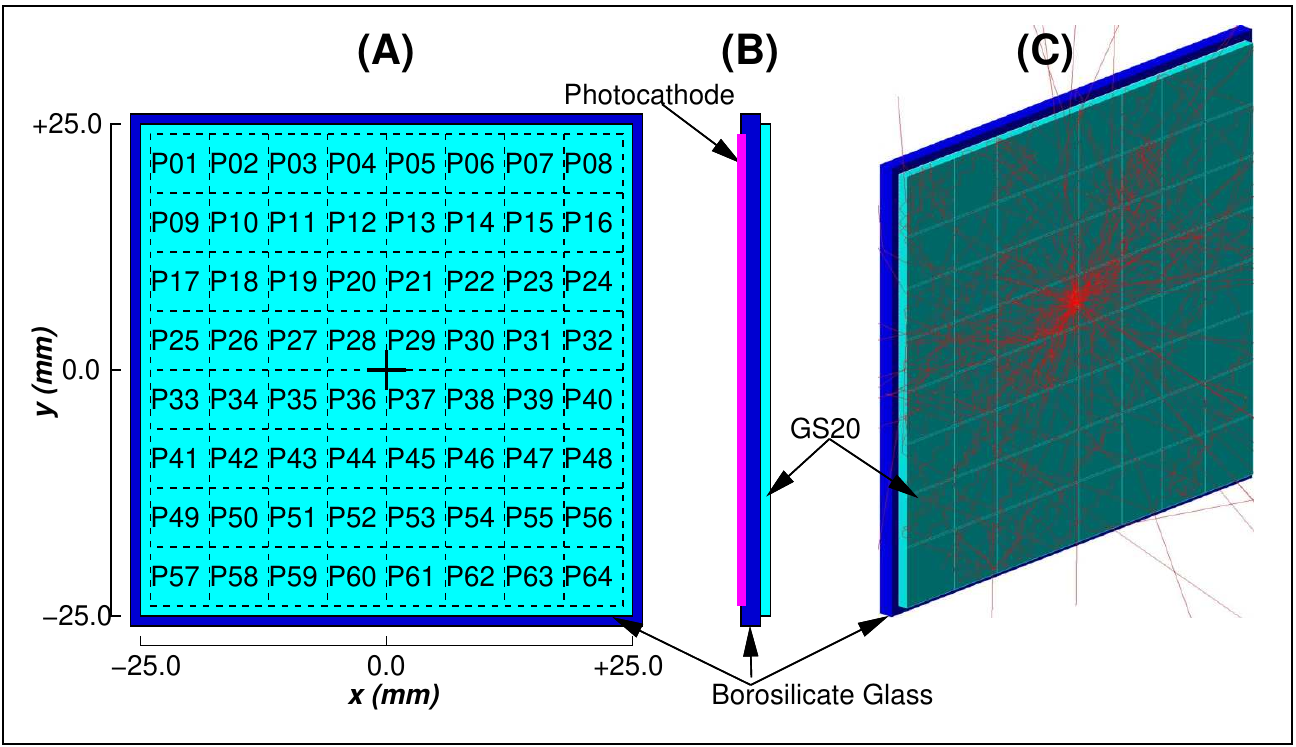}.

\caption{\label{fig:G4}Rendering of the Geant4 model of a SoNDe module. (A)
Front view where the cross denotes the x-y coordinate origin and the
dashed lines show the photocathode pixel boundaries. The z origin
is at the centre of the GS20 sheet. (B) Side view showing the photocathode
(magenta line) attached to the rear face of the glass window of the
MAPMT. (C) 3D view where the red tracks show the scattering of 100
optical photons in the glass components.}
\end{figure}

A rendering of the geometry encoded in the computer model is displayed
in Fig.~\ref{fig:G4}. This shows the scintillator, with optional
grooving etched into the surface of the glass. Also displayed is the
borosilicate glass of the MAPMT window and the tracks of optical photons
(red lines, Fig.~\ref{fig:G4}(C)) started at the mid point of pixel
P28. The origin of the coordinate system, shown by the cross in Fig.~\ref{fig:G4}A,
is the centre of the cuboid volume which defines the GS20 sheet.

The dimensions of the GS20 sheet have been set to $50.0\times50.0\times1.0$~$\mathrm{mm^{3}}$
to match the size of the piece used in test measurements. A thickness
of 1~mm produces a thermal-neutron detection efficiency of $\sim76$\%.
The effectiveness of dividing the GS20 sheets into pixels to reduce
the spread of signal was investigated. Grooves in the glass surface,
following the pixel boundaries of the MAPMT with variable depth and
width, could be introduced optionally. The default surface finish
of the GS20 sheet was polished and optional polished Al or matt TiO$_{2}$
reflector were included at the external front face of the sheet.

The dimensions of the H12700A MAPMT were taken from Ref.~\citep{MAPMT}.
Only the glass window, dimensions $52.0\times52.0\times1.5$~$\mathrm{mm^{3}}$,
and the bialkalai photocathode, dimensions $48.5\times48.5$~$\mathrm{mm^{2}}$,
deposited on the back side of the window are modelled. Metalic side
walls and internal electrode structures were not included in the model.

\begin{figure}[h]
\includegraphics[width=1\columnwidth]{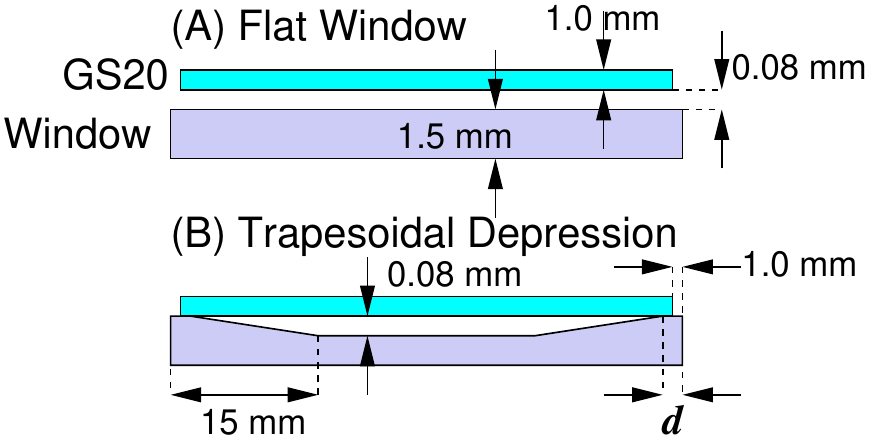}

\caption{\label{fig:MAPMT-window-detail}MAPMT window detail (not to scale).}

\end{figure}

Although nominally flat, the borosilicate glass window of the MAPMT
surface is slightly depressed at the centre with respect to the edge.
Measurements were made of 35 MAPMT windows, yielding a mean depression
of 0.08 mm with a standard deviation \textgreek{sv} = 0.02 mm. The
exact shape of the window surface was not determined, but where there
was no coupling medium between GS20 and MAPMT in the simulation, a
small air gap of thickness 0.08 mm was introduced between the GS20
and the borosilicate glass window of the  MAPMT (Fig.~\ref{fig:MAPMT-window-detail}(A)).
Where a coupling medium (e.g. optical epoxy) was introduced it also
had a thickness of 0.08~mm. Most of the calculations described in
Sec.~\ref{sec:Optical-Transport} have been made with a constant-thickness
air gap. However the effect of a non-flat MAPMT was modelled by introducing
a trapesoidal depression (Fig.~\ref{fig:MAPMT-window-detail}(B))
to the window surface. Comparisons of flat and non-flat windows are
given in Sec.~\ref{subsec:non-flat}.

\subsection{\label{subsec:Optical-Properties}Optical Properties}

Optical photons were tracked \citep{G4Phys} through the glass components
of the detector as they undergo Rayleigh scattering, absorption and
medium-boundary processes. Boundary processes, which were simulated
using the `glisur model' of Geant4, include:
\begin{enumerate}
\item dielectric to dielectric where the photon is refracted, ie transmitted
through the boundary or reflected at the boundary.
\item dielectric to metal where the photon is absorbed in the `metal'
or reflected at the boundary.
\item dielectric to black where the photon is absorbed at the boundary.
\end{enumerate}
The surfaces of the GS20, borosilicate PMT glass and any optional
coupling medium were assumed to be polished, so that reflection was
specular, but the effect of small surface irregularities was also
calculated via the `polish' parameter incorporated in the glisur
model \citep{G4main,Polish-fact}. This is also investigated in Sec.~\ref{subsec:Monoenergetic Protons}.

\begin{figure}[h]
\begin{center}\includegraphics[width=0.75\columnwidth]{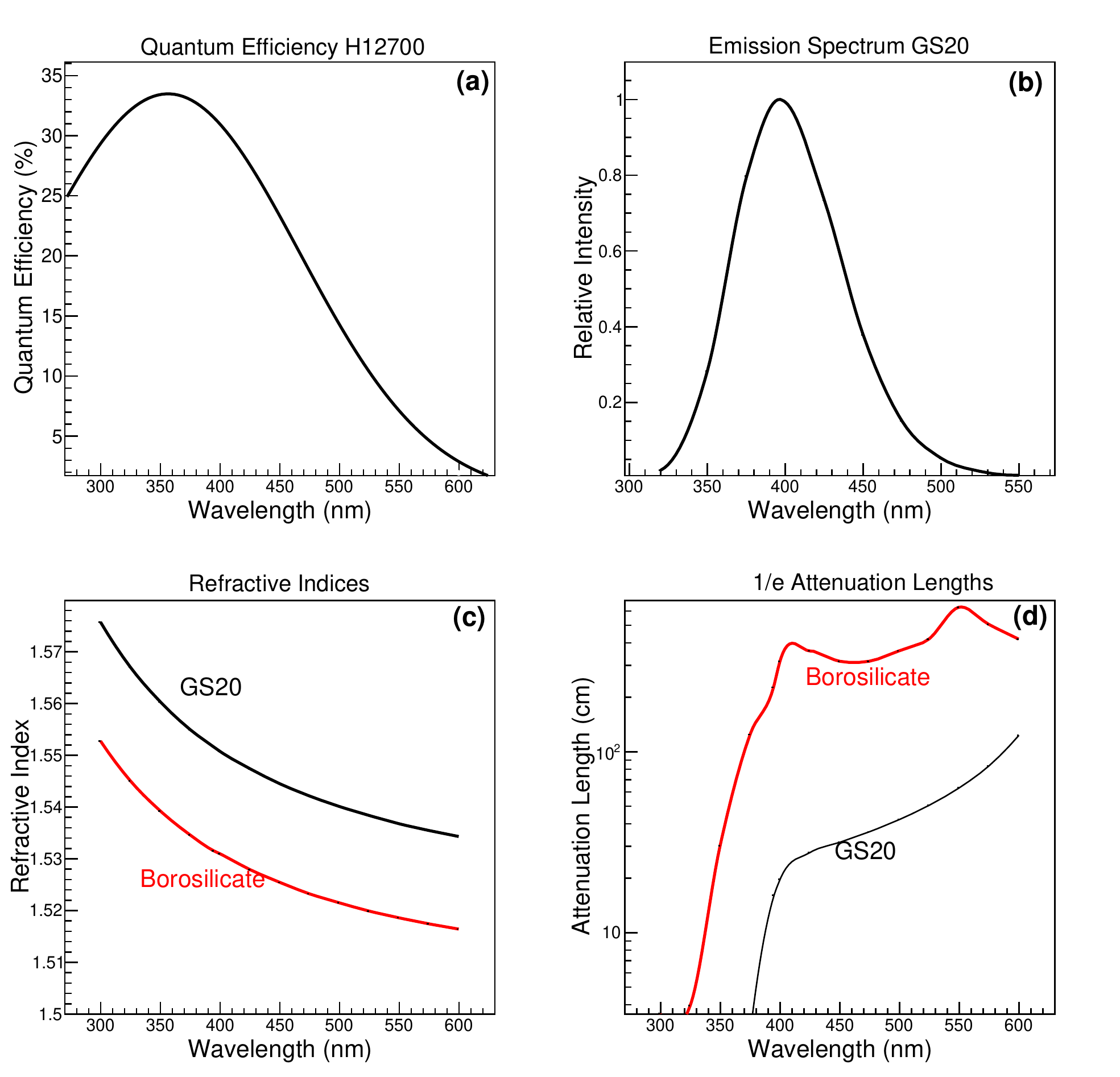}\end{center}

\caption{\label{fig:Optical-properties}Optical properties employed in the
simulation: (a) the quantum efficiency of the bialkalai cathode of
the H12700 PMT \citep{MAPMT}, (b) the scintillation emission spectrum
of GS20 \citep{GS20-1}, (c) refractive indices of GS20 \citep{GS20-1}
and borosilicate glass \citep{borosilicate}, (d) $1/e$ attenuation
lengths of GS20 \citep{GS20-1} and borosilicate glass \citep{borosilicate}.}
\end{figure}

The optical properties of each material such as refractive index,
$1/e$ attenuation length and reflectivity have been entered, for
photon energies in the range 2.066~eV (600~nm) to 4.133~eV (300~nm),
into the properties tables linked to particular materials. Some of
the employed optical parameters are displayed in Fig.~\ref{fig:Optical-properties}.
EJ500 \citep{EJ-500}, the optional optical-coupling epoxy, was given
a constant refractive index of 1.57.

A light output of 20\% of anthracene (3500 photons/MeV), and a Birks
parameter \citep{Birks} $k_{B}=0.01\:\mathrm{mm/MeV}$ has been used
for GS20 scintillator. The Birks parameter models the non-linear (with
respect to energy deposit) response of the scintillator, which is
due to increased quenching of the scintillation signal when $dE/dx$
and the ionisation density along a charged track are large (see Eq.~\ref{eq:2_birks}).
The choice of Birks parameter value is discussed in Sec. \ref{subsec:Comparison-with-proton}.

The optional outer reflector scintillator coatings were either matt
$\mathrm{TiO_{2}}$ paint \citep{EJ-520} or polished Al foil. The
former has $\sim91$\% reflectance at 390~nm, falling to 28\% at
350~nm, while the latter has a reflectance of 92\% at 390~nm, with
a small wavelength dependence \citep{Al-refl}.

\subsection{\label{subsec:Modes-of-operation}Modes of operation}

The code may be run in three different modes which are described below.
In all cases information has been recorded event-by-event in ROOT
TTree format \citep{ROOT} and then analysed using a ROOT C++ macro.

\subsubsection{\label{subsec:Optical-Photon-Transport}Optical Photon Transport}

Optical photons were generated isotropically at a wavelength of 395~nm
(3.15 ~eV), at a chosen point inside the GS20 sheet, using the General
Particle Source tool of Geant4. The wavelength corresponds to the
peak of the scintillation emission spectrum of GS20. When a photon
impinged on the photocathode, where it was absorbed, the photon wavelength,
direction and interaction position were recorded. Otherwise if absorbed
outside the photocathode or scattered outside of the detector volume
the photon was lost. 

\subsubsection{\label{subsec:Interactions-of-Ionising}Interactions of Ionising
Radiation in GS20}

The interactions of ionisating radiation within the sheet of GS20
have been modelled using the high-precision hadronic interaction class
FTFP\_Bert\_HP and electromagnetic interaction classes G4EmStandardPhysics
and G4EmExtraPhysics. For incident neutron energies below 4.0~eV
the interaction class \\
G4NeutronHPThermalScattering with data base \\
G4NeutronHPThermalScatteringData was employed.

Primary and any subsequent  (from gamma rays or neutrons) charged
particles were tracked and the energy losses through ionisation recorded.
At each discrete step along a charged-particle track the energy loss
along the step, the time with respect to particle production at source
and the mean step position were recorded in the ROOT TTree.

\subsubsection{\label{subsec:Full-Calculation}Full Calculation}

This mode proceeded as in Sec.~\ref{subsec:Interactions-of-Ionising},
but included the production of scintillation photons along the charged-particle
tracks. Cherenkov radiation was also produced if the charged-particle
velocity was above threshold for a particular optical medium. The
optical photons were then tracked as in Sec.~\ref{subsec:Optical-Photon-Transport}.
In this mode, information for both the charged-particle track in the
GS20 sheet and the optical photon impinging on the photocathode was
recorded.

\section{\label{sec:Optical-Transport}Optical Photon Transport Simulations}

\begin{figure}[h]
\begin{center}\includegraphics[width=0.9\columnwidth]{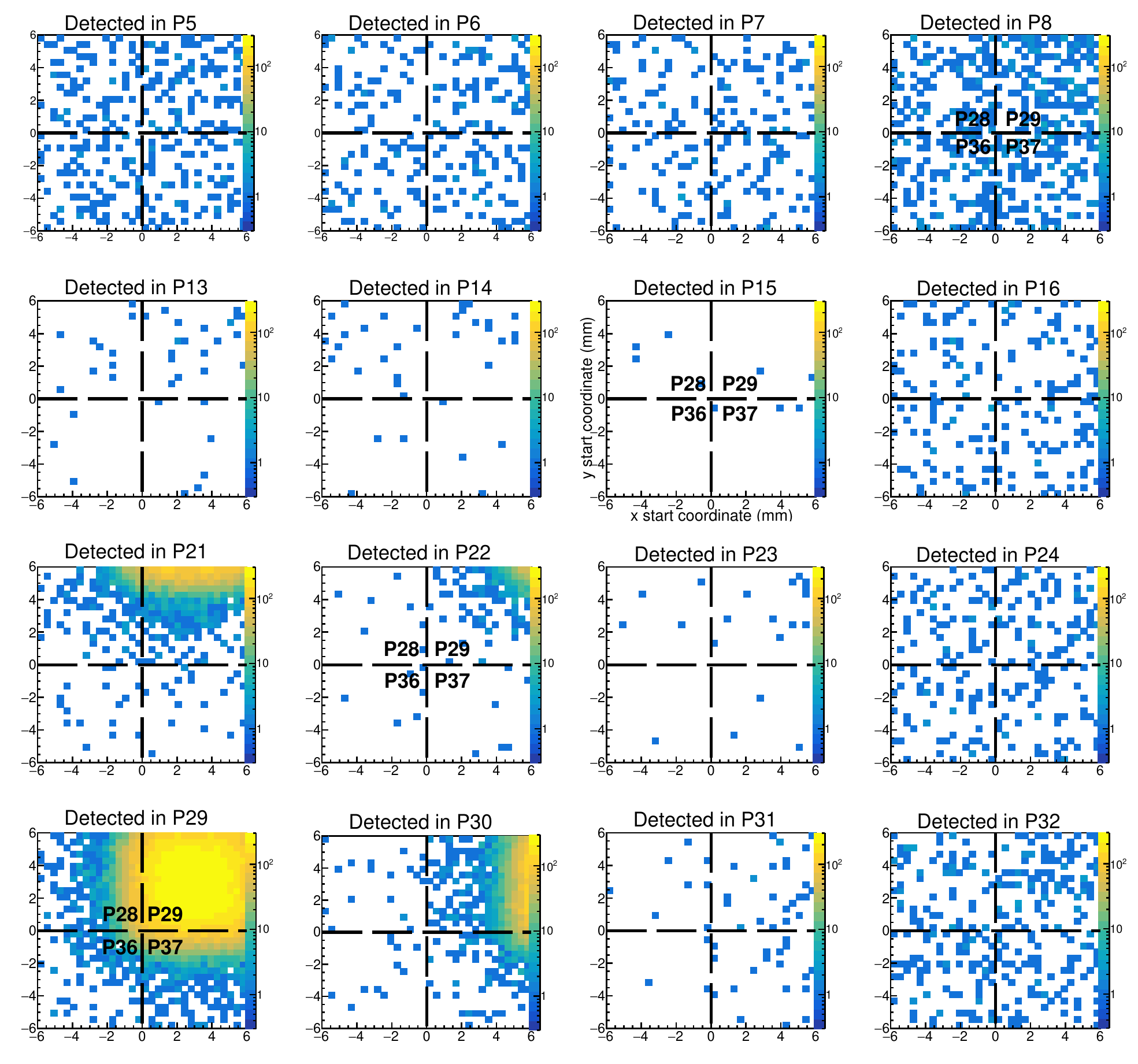}\end{center}

\caption{\label{fig:Photon-spread}}
Starting $(x,y)$ coordinates for photons registered in the top-right
quadrant of MAPMT pixels. Dashed lines show pixel boundaries between
P28, P29, P36 and P37. Pixel numbering is given in Fig.~\ref{fig:G4}.
Note that the z colour scale is logarithmic.
\end{figure}

The transport of optical photons was simulated, for a variety of configurations
of the SoNDe module, to calculate their spreading away from the point
of origin before detection at the MAPMT photocathode. Parameters such
as the grooving, optical coupling between scintillator and borosilicate
glass, and reflective coatings for the scintillator were varied in
various combinations. The optical photons were started at a random
$x-y-z$ position within the GS20 and assigned a random direction
of polarisation. 

The 1 mm thick Li-glass sheet in front of the 4 pixels (P28, P29,
P36, P37) bounding the centre of the MAPMT was seeded uniformly with
scintillation photons. The volume seeded was a cuboid centred at coordinate
(0,0,0) (see cross in Fig. \ref{fig:G4}(A)) and the photon starting
coordinates were constrainted to be within $-6<x<+6$~mm, $-6<y<+6$~mm,
and $-0.5<z<+0.5$~mm. The configuration of the SoNDe module was
as in Run $R_{11}$ (introduced and described below). Each scintillation
photon with a unique, randomly-generated $(x,y,z)$ start coordinate
was then tracked until it struck a photocathode pixel, where its initial
and final coordinates were recorded. Fig.~\ref{fig:Photon-spread}
shows the results of this tracking, illustrating the correlation between
the start coordinate of a photon and the pixel where the photon was
finally detected. Each panel displays events where the photon has
been detected in the corresponding photocathode pixel. The panels
have been populated with the $(x,y)$ start coordinates of the photons
which are detected in the corresponding pixels. The 16 pixels P5 --
8, P13 -- 16, P21 -- 24 and P29 -- 32 correspond to the top right-hand
$4\times4$ quadrant of the MAPMT (Fig.~\ref{fig:G4}(A))

Evidently, hits in P29 were overwhelmingly due to photons seeded in
front of P29, but a significant number of photons seeded in front
of P28, P36 and P37, mostly within $\sim2$~mm of the boundary to
P29, were also registered in P29. Similarly for hits in the nearest
neighbours to P29: P21 (vertical), P30 (horizontal) and P22 (diagonal),
the start coordinates are clustered close to the neighbour's boundary
with P29. MAPMT edge pixels P5 -- 8, P16, P24 and P32 record more
photons than P13 -- 15, P23 and P31, which are closer to the photon
source, due to reflections at the thin side edge of the GS20 sheet.
In these edge pixels, there is no obvious correlation between the
start position and the pixel where the photon is detected. 

\begin{figure}[h]
\includegraphics[width=1\columnwidth]{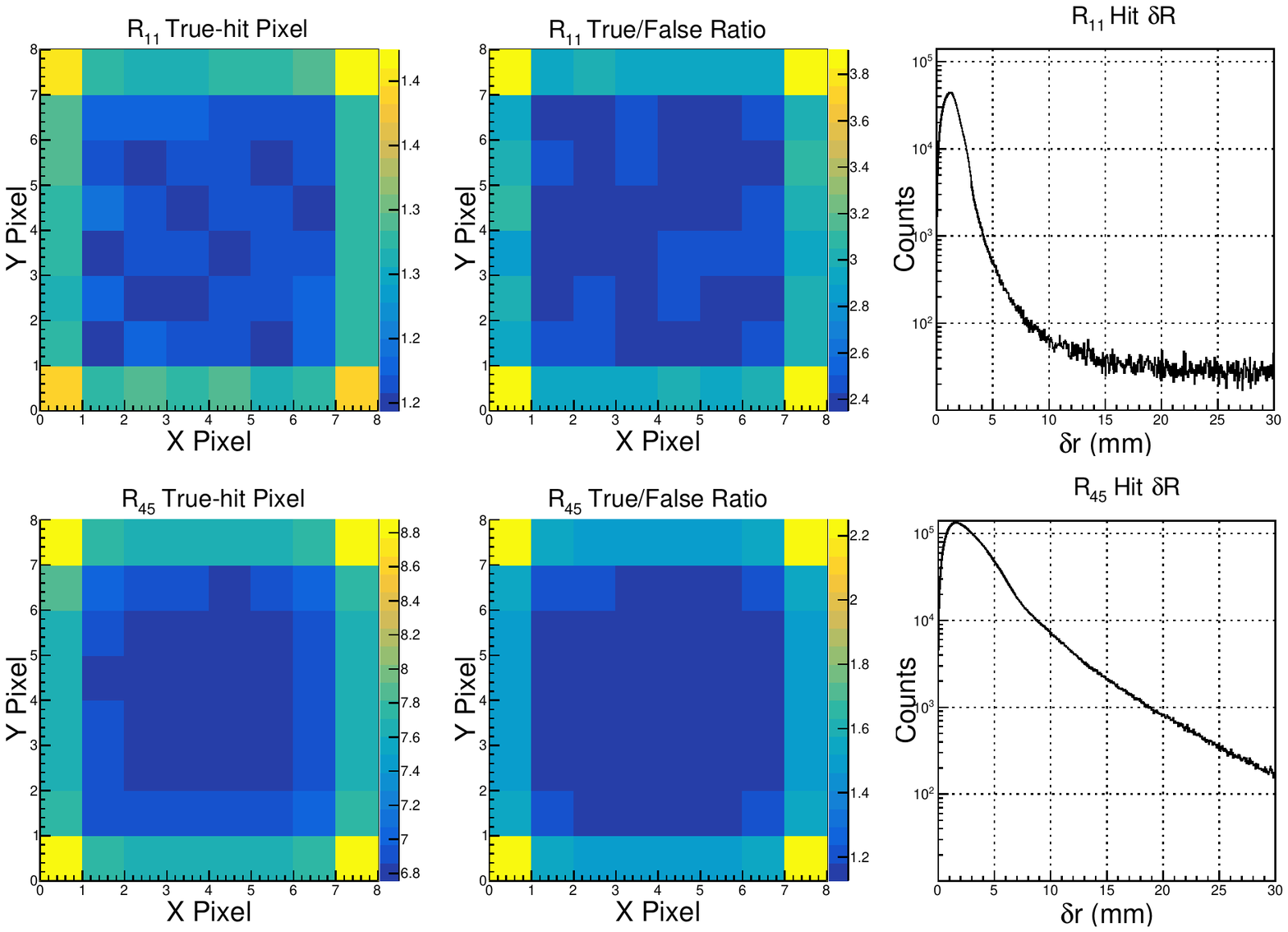}

\caption{\label{fig:Run-11}Top row R$_{11}$ results, left values of $\epsilon_{T}$
for each illuminated pixel, centre values of $R_{TF}$ for each illuminated
pixel, right $\delta r$ for entire illuminated region. Bottom row
R$_{45}$ results, order as in top row. Note that the colour vertical
scales are different for each 2D pixel plot.}
\end{figure}

\begin{figure}[h]
\begin{center}\includegraphics[width=1\columnwidth]{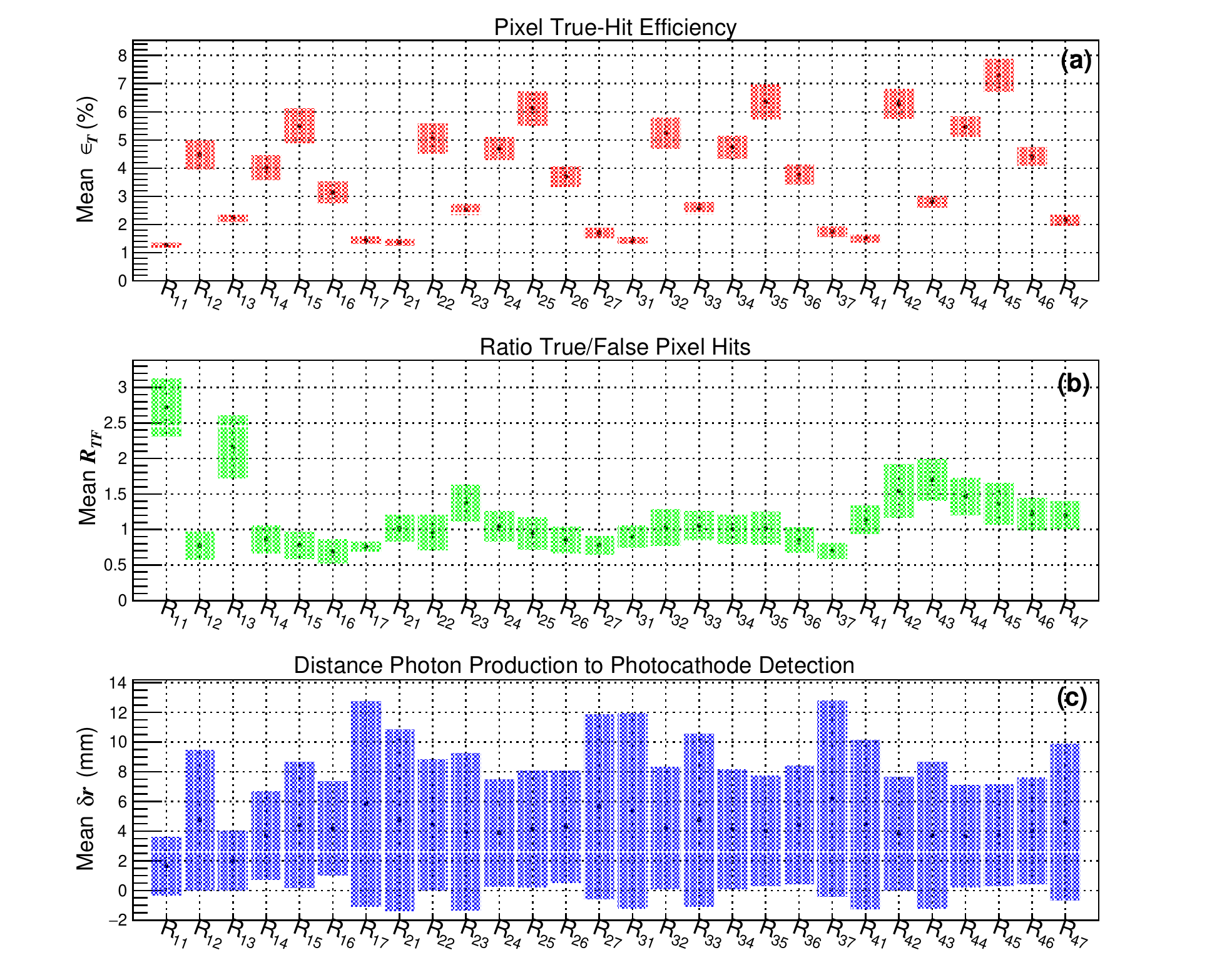}\end{center}

\caption{\label{fig:photon-summary}Summary of photon transport results: a)
mean values $\epsilon_{T}$; b) mean values $R_{TF}$; c) mean values
$\delta r.$ The x-axis labels refer to the run labels R$_{ij}$ explained
in the text. The vertical extents of the filled boxes, display the
standard deviations (Eq.~\ref{eq:std}).}
\end{figure}

Further runs were made with the $x-y$ start coordinates covering
the entire GS20 sheet: $-25<x<+25$~mm, $-25<y<+25$~mm and again
the $z$ coordinate was constrained to be within the 1~mm thickness.
The simulation was split into 28 runs, each starting $8\times10^{7}$
optical photons, with direction sampled from an isotropic angular
distribution. They were arranged in 4 groups of 7. The 4 groups corresponded
to conditions:
\begin{enumerate}
\item No grooving of the GS20 sheet.
\item Grooving 0.5~mm deep and 0.2~mm wide, cut on the outside face of
the GS20 sheet.
\item Grooving 0.5~mm deep and 0.2~mm wide, cut on the inside face of
the GS20 sheet.
\item Grooving 1.0~mm deep (i.e. all the way through the sheet) and 0.2~mm
wide.
\end{enumerate}
Within each group, the 7 sub-groups correspond to conditions:
\begin{enumerate}
\item No external reflector on the outside face of the GS20 sheet and no
optical coupling between the GS20 and MAPMT window.
\item TiO$_{2}$ reflector on the outside face of the GS20 and no optical
coupling.
\item Polished Al reflector on the outside face of the GS20 and no optical
coupling.
\item Polished Al reflector on the outside face of the GS20 and optical
epoxy coupling between the GS20 and MAPMT window
\item TiO$_{2}$ reflector on the outside face of the GS20 and optical epoxy
coupling.
\item No reflector and optical epoxy coupling.
\item As sub-group  1, but with the surface polish factor for the GS20 sheet
set to $P=0.8$. $P$ values may be varied from 0.0 to 1.0 \citep{Polish-fact}
where the latter denotes a perfect surface. The MAPMT surface was
set to perfectly polished ($P=1.0$).
\end{enumerate}
Runs have been labelled R$_{ij}$, where $i=1,4$ is the group and
$j=1,7$ is the sub-group. Analysis of the Monte Carlo output has
determined the $x-y$ translation distance, $\delta r=\sqrt{\delta x^{2}+\delta y^{2}}$,
where $\delta x$ and $\delta y$ are the differences between the
photon-start coordinates and the photon-detection coordinates at the
MAPMT photocathode. The MAPMT hardware can resolve detected photon
position down to the $6\times6$~mm photocathode pixel size. Thus
the analysis determines if the photon-start pixel is the same as the
photocathode-detection pixel, which is denoted a true hit. Where the
start and detection pixels are not the same, this is denoted a false
hit. If $n_{T}$ and $n_{F}$ denote the numbers of true and false
hits respectively within a given pixel, the true-hit efficiency $\epsilon_{T}=100\times n_{T}/(n_{T}+n_{F})$
is just the ratio of detected-to-started photons within a given pixel
(expressed in \%), while $R_{TF}=n_{T}/n_{F}$ is the ratio of the
number of true-to-false hits in a given pixel.

Representative distributions are given for runs R$_{11}$ and R$_{45}$
in Fig. \ref{fig:Run-11}, which displays the pixel distributions
of $\epsilon_{T}$ and $R_{TF}$ and $\delta r$ for the entire array.
While the $\epsilon_{T}$ are much larger for R$_{45}$, where the
TiO$_{2}$ reflector increases the overall light-collection efficiency
significantly, the $R_{TF}$ are significantly higher for R$_{11}$
(no reflector) which correlates with a significantly smaller average
$\delta r$.

A summary of the results for all runs is displayed in Fig.~\ref{fig:photon-summary}.
Fig.~\ref{fig:photon-summary}(a) displays $\overline{\epsilon}_{T}$
the mean value of the $\epsilon_{T}$ averaged over all illuminated
pixels. The error-bars show the standard deviations of the individual-pixel,
true-hit efficiencies

\begin{equation}
\sigma_{\epsilon}=\sqrt{\frac{\stackrel[{\scriptscriptstyle i=1}]{{\scriptscriptstyle N}}{\sum}(\epsilon_{i}-\overline{\epsilon}_{T})^{2}}{N-1}}\label{eq:std}
\end{equation}
 where $\epsilon_{i}$ are the individual-pixel, true-hit efficiencies
and $N$ is the number of illuminated pixels. Fig.~\ref{fig:photon-summary}(b)
displays $\overline{R}_{TF}$ the pixel-averaged mean of the $R_{TF}$
for each run, and again the error bars denote the standard deviations
of the individual-pixel ratios. Fig.~\ref{fig:photon-summary}(c)
displays $\overline{\delta r}$ the mean value (averaged over all
illuminated pixels) of $\delta r$ for each run, where the error bars
denote the standard deviation of the mean, related to the spread in
event-by-event $\delta r$ values over the entire illuminated region.

Comparing the main group runs, grooving of the GS20 sheet does increase
$\overline{\epsilon}_{T}$ slightly, but compared to no grooving $\overline{R}_{TF}$
is smaller and $\overline{\delta r}$ is slightly larger on average.
Within the sub-groups, TiO$_{2}$ matt reflector and optical epoxy
coupling produce the highest $\overline{\epsilon}_{T}.$ However a
polished Al reflector gives better $\overline{R}_{TF}$ especially
if no epoxy coupling is employed. Overall, the best $\overline{R}_{TF}$
is obtained in R$_{11}$: no grooves, no reflector and no epoxy coupling,
although in this case $\overline{\epsilon}_{T}$ is the lowest of
all runs. This results from the mismatch of refractive indices at
the GS20-air-borosilicate glass boundary, which effectively selects
for transmission only photons close to perpendicular incidence, reducing
the spread of detected photons from their point of origin. Thus $\overline{\epsilon}_{T}$
is reduced, while $\overline{R}_{TF}$ is increased.

Comparing $R_{11}$ ($P=1.0)$ and $R_{17}$ ($P=0.8$), the latter
shows a large reduction in $\overline{R}_{TF}$, a large increase
in $\overline{\delta r}$ and a small increase in $\overline{\epsilon}_{T}$. 

\subsection{\label{subsec:non-flat}Comparison of flat and non-flat MAPMT window}

\begin{figure}[h]
\includegraphics[width=1\columnwidth]{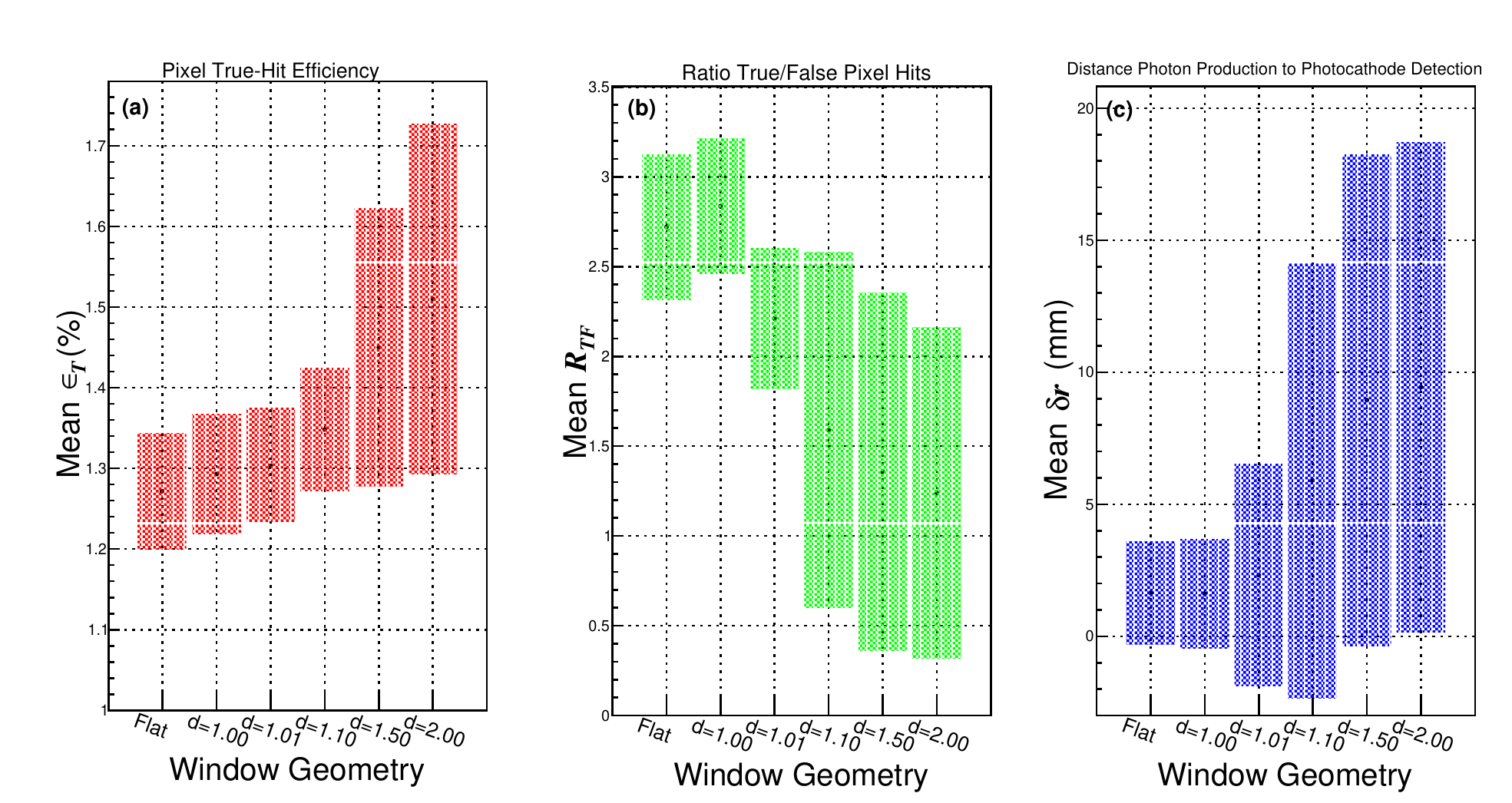}

\caption{\label{fig:Window-geometry-dependence}Window-geometry dependence:
(a) true-hit efficiency $\epsilon_{T}$; (b) true-hit/false-hit ratio
$R_{TF}$; (c) $\delta$R; The vertical extents of the filled boxes,
display the standard deviations (Eq.~\ref{eq:std}) of the respective
parameters.}
\end{figure}

Optical simulations were performed for 6 geometries of the MAPMT window
(Fig.~\ref{fig:MAPMT-window-detail}): (1) flat face; (2) trapesoidal
face $d=1.00$~mm; (3) trapesoidal face $d=1.01$~mm; (4) trapesoidal
face $d=1.10$~mm; (5) trapesoidal face $d=1.50$~mm; (6) trapesoidal
face $d=2.00$~mm; where the width of the overlap $(d-1.0\:\mathrm{mm)}$
defines the area of GS20-glass contact at the edge of the sheet. Otherwise
the SoNDe module was in $R_{11}$ configuration. The results are summerised
in Fig.~\ref{fig:Window-geometry-dependence}. Cases 1 and 2 are
very similar, so that the shape of the air gap, trapesoidal or constant
thickness, makes little difference. Where parameter $d>1.0$~mm (cases
3 - 6) , there is an area of contact between the GS20 and borosilicate
window. As $d$ increases $\overline{\epsilon}_{T}$ increases slightly,
$\overline{R}_{TF}$ decreases and $\overline{\delta r}$ increases.
The spread of values of these parameters also increases with $d$,
as the divergence in response between edge pixels, close to the contact
area, and more central pixels increases.

The exact shape of the MAPMT surface has not been measured, but it
seems unlikely that there will be a large contact area between the
GS20 and borosilicate window. Subsequent calculations in Sec.~\ref{sec:Particle-response}
have been made with a flat MAPMT window and correspond to run case
$R_{11}$ described above.

\section{\label{sec:Particle-response}Response to Ionising Radiation}

GS20 is designed to detect thermal neutrons but scintillating glass
is sensitive to ionising radiation in general. The 1~mm thickness
of the GS20 sheet is a compromise between high detection efficiency
($\sim76\%$, Sec.~\ref{subsec:Thermal-and-Fast}) for thermal neutrons
and low detection efficiency for background radiation. Increasing
the thickness will increase the background efficiency preferentially,
relative to that of thermal neutrons. In operation at ESS the potential
sources of background will be gamma rays \citep{gamSens} and fast
neutrons \citep{fnSens}, which will range in energy from sub-MeV
to around GeV.

When comparing the response of a scintillator to low-energy particles
of different types, non-linear effects due to quenching of the scintillation
process must be taken into consideration. This is performed in Geant4
using the empirical formula of Birks \citep{Birks}

\begin{equation}
\frac{dL}{dx}=S\frac{\frac{dE}{dx}}{1+k_{B}\frac{dE}{dx}}\label{eq:2_birks}
\end{equation}

which relates the scintillation light yield per unit path $dL/dx$
to the scintillation efficiency $S$, the differential charged-particle
energy loss $dE/dx$ and a material-dependent constant $k_{B}$, often
known as the Birks parameter. This parameter is not well established
for the glass GS20. A study of a variety of inorganic crystaline scintillators
\citep{Tretyak} has produced values of $k_{B}$ in the range $10^{-3}-10^{-2}\:\mathrm{g/cm^{2}/MeV}$.
A value: $k_{B}=10^{-2}\:\mathrm{mm/MeV}$ has been employed for GS20
(equivalent to $2.5\times10^{-3}\:\mathrm{g/cm^{2}/MeV})$, which
is discussed in Sec.~\ref{subsec:Comparison-with-proton}.

\subsection{\label{subsec:Alpha-Particle-Response}Position-Dependent Alpha-Particle
Response}

\begin{figure}[h]
\includegraphics[width=1\columnwidth]{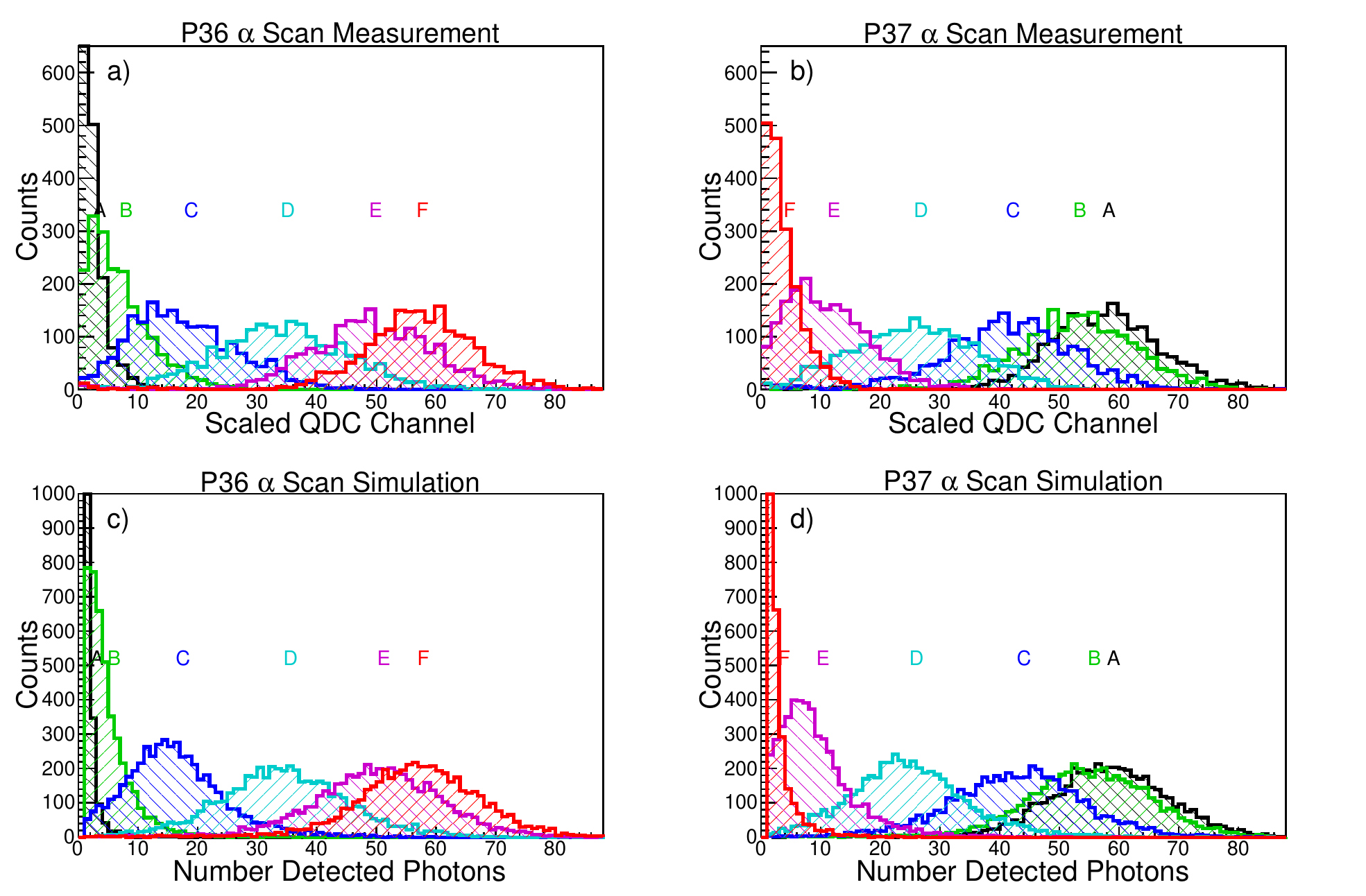}

\caption{\label{fig:Alpha-horizontal-scan} Alpha source horizontal scan signal
amplitude spectra: a) P36 measurement; b) P37 measurement; c) P36
simulation; d) P37 simulation.}
\end{figure}

\begin{figure}[h]
\includegraphics[width=1\columnwidth]{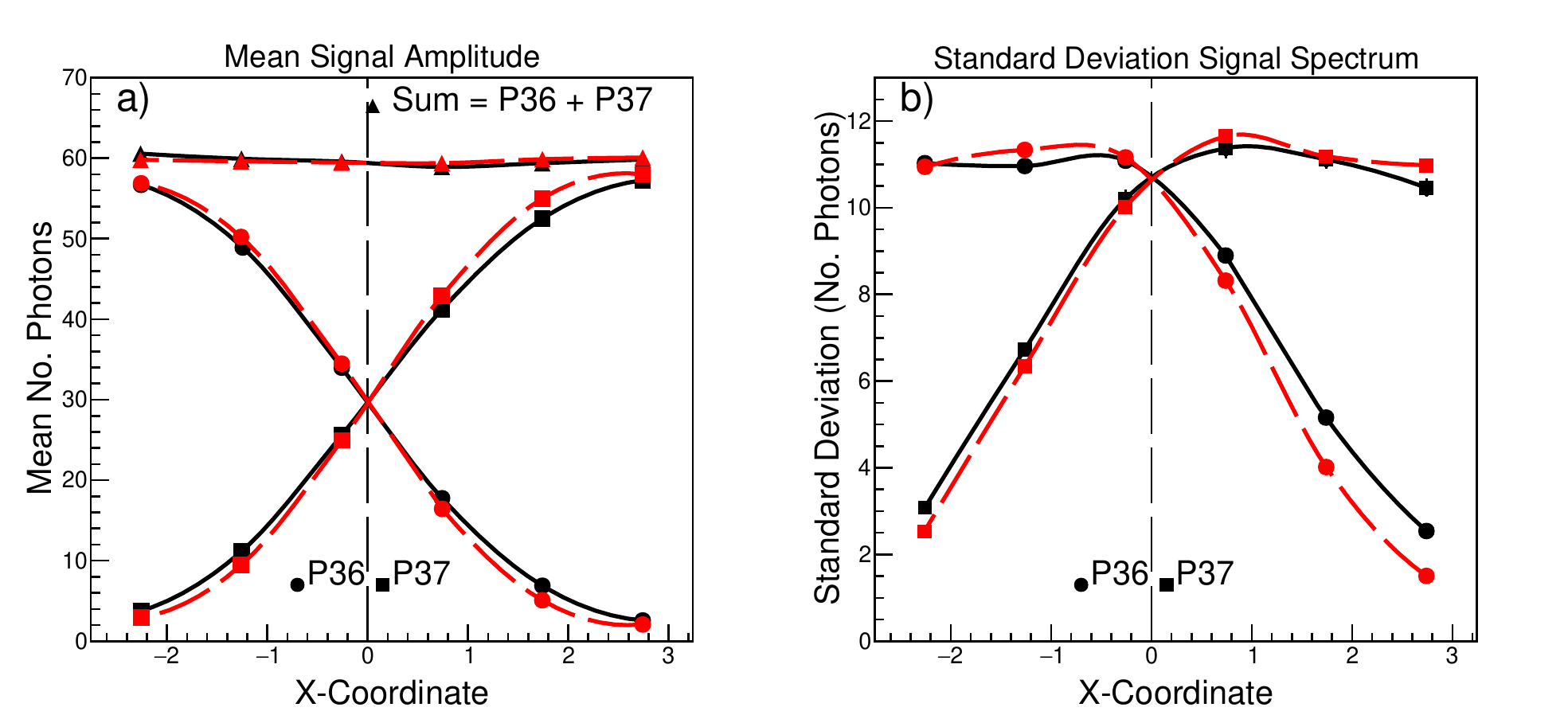}

\caption{\label{fig:Alpha-mean-vs}Alpha source scan position dependence: a)
measured (black) and simulated (red) signal amplitudes; b) measured
(black) and simulated (red) standard deviations of signal amplitude
spectra. Circles denote pixel P36 and squares P37. The dashed line
denotes the vertical boundary between P36 and P37.}
\end{figure}

A collimated $^{241}\mathrm{Am}$ alpha-particle source was employed
to measure the position dependence of the signal amplitude from the
individual MAPMT pixels of a SoNDe module. Measurements of the alpha
response, as the collimated source was scanned across the face of
the GS20 sheet, have been reported in Ref.~\citep{Rofors-Alpha}
and these measurements have been simulated with the present computer
model. Incident alpha-particle energies were sampled from the measured
energy spectrum of the employed $^{241}\mathrm{Am}$ source and their
direction was restricted to a cone of half-angle $6.2^{\circ}$ and
axis normal to the plane of the scintillator sheet, to reproduce the
experimental collimation. The alpha particles passed through $\sim6$~mm
of air before incidence on the GS20, so that their mean energy at
the scintillator was $\sim4$~MeV (see Fig.~2 of Ref.~\citep{Rofors-Alpha})
and the diameter of the illuminated spot at the GS20 face was $\sim1.3$~mm.
There was $\sim0.2$~mm uncertainty in the experimental beam spot
diameter due to uncertainties in the internal dimensions of the metal
source container. 

Fig.~\ref{fig:Alpha-horizontal-scan} compares the measured alpha-particle
spectra, from MAPMT pixels 36 and 37 (Fig.~\ref{fig:G4}(A)), to
the corresponding simulated spectra for a horizontal scan over positions
labeled A-F (see Fig.~5 of Ref.\citep{Rofors-Alpha}). Labels A,
B, C, D, E, F correspond to \emph{x} coordinates 2.74, 1.74, 0.74,
-0.26, -1.26, -2.26~mm respectively, with the \emph{y} coordinate
fixed at -3.4~mm. The coordinate system is displayed in Fig.~\ref{fig:G4}(A).
Fig.~\ref{fig:Alpha-mean-vs} compares the position dependence of
the mean values and widths of the spectral distributions at the 6
scanned \emph{x} coordinates. The bin widths of the measured histograms
(QDC channel in Ref.~\citep{Rofors-Alpha}) have been scaled, so
that the average of the 6 summed (P36 + P37) points is equal to the
corresponding simulated quantity, i.e. the measured mean signal charge
has been normalised to the calculated mean number of detected scintillation
photons.

The agreement between the measured and simulated position dependence
of the mean pulse height is excellent. The agreement for the position
dependence of the widths (standard deviations of the spectral distributions)
is also good, although the simulation slightly overpredicts the change
in signal amplitude as one moves from position A to F. This would
be expected if effects such as electronic noise, which were not simulated,
made a non-negligible contribution to the measurement. This is explored
further in Sec.~\ref{subsec:Monoenergetic Protons}.

\subsection{\label{subsec:Monoenergetic Protons}Position-Dependent Response
to Monoenergetic Protons}

\begin{figure}[h]
\includegraphics[width=1\columnwidth]{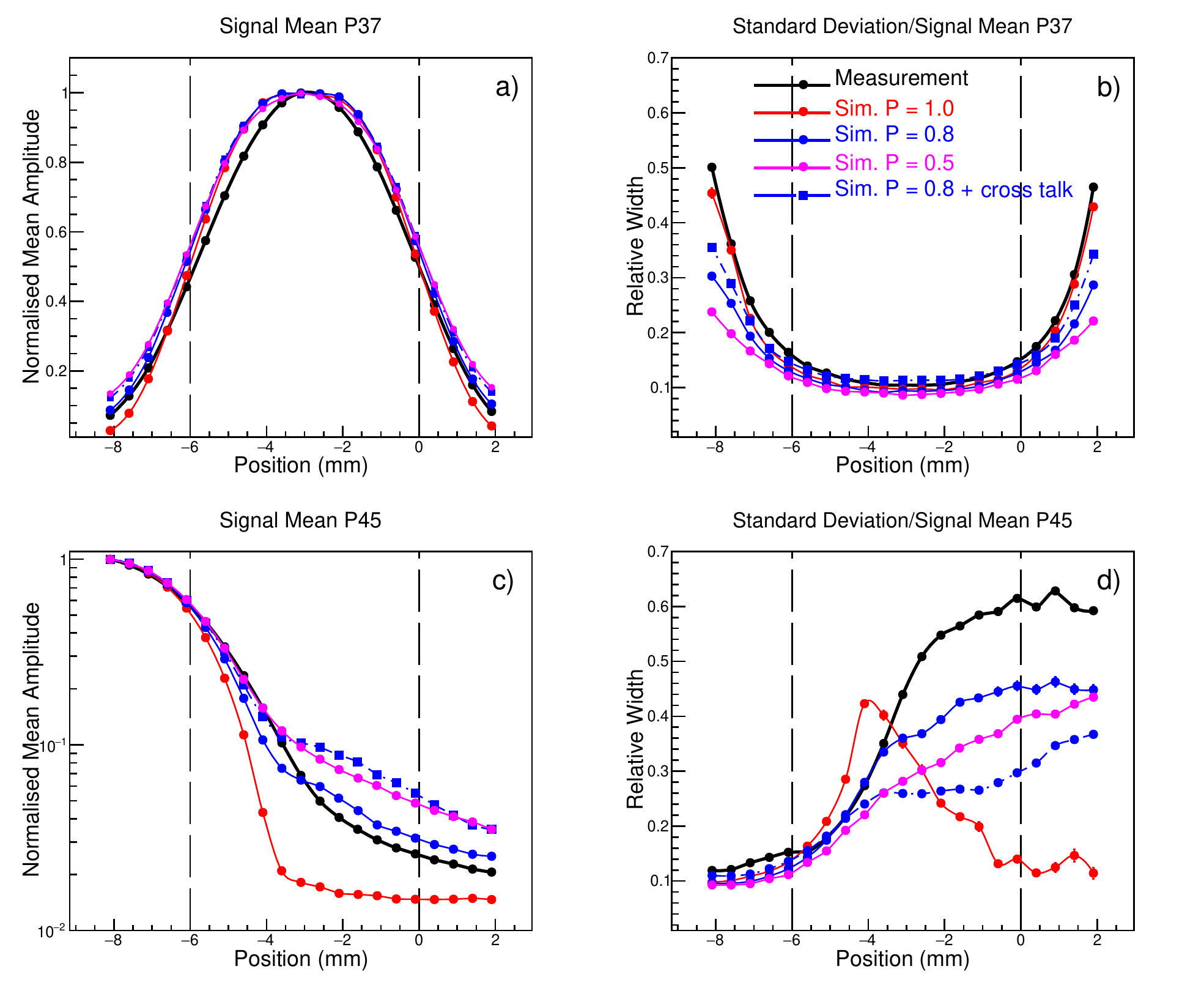}

\caption{\label{fig:Vertical-proton-scan.}Vertical proton scan: means and
standard deviations of pulse-height spectra from P37 (plots (a), (b))
and P45 (plots (c), (d)). Colour coding is shown in plot (b) and applies
to all sub-plots. The vertical dashed lines show the boundaries of
pixels 45-37 (left) and 37-29 (right).}

\end{figure}

The Lund LIBAF facility \citep{LIBAF} provides finely-collimated,
monoenergetic beams of protons and deuterons at energies up to 3~MeV.
Ref. \citep{Rofors_proton} describes detailed scans of LIBAF proton
beams across the face of a SoNDe module and gives a first comparison
of the proton measurements with the simulation. Here we show a further
comparison with unpublished data taken at LIBAF and the latest version
of the simulation.

Two linear scans were made: horizontal covering coordinates 2.4 -
12.9~mm at fixed $y=1.90$~mm, in 0.5 mm steps and vertical covering
coordinates -5.6 - 1.9~mm at fixed $x=8.4$~mm, in 0.5~mm steps.
The beam energy was set to 2.5~MeV and energy losses in the beam
line and air before protons hit the GS20 sheet amounted to around
20~keV. Details of the proton beam line and x-y scanner are found
in Ref.~\citep{Rofors_proton}. The MAPMT was operated at a high
voltage of -1.0~kV and data were collected using a VMEbus-based,
data-acquisition system similar to that described in Ref.~\citep{VME-daq}
with the pulse-height spectra accumulated in CAEN V792 charge-to--digital
converters.

Fig.~\ref{fig:Vertical-proton-scan.} compares the vertical-scan
measurements (black circles and lines), for vertically adjacent pixels
P37 and P45, with equivalent simulations. Horizontal scan data produced
very similar distributions. The base-level simulation (red circles
and lines), equivalent to case $R_{11}$ above, modeled optical transport
and did not consider electronic effects. A perfectly smooth surface
was assumed initially for the GS20 and MAPMT glass, which is parametrised
by a `polish' factor $P=1.0$. The effect of a non-perfect GS20
surface was then investigated with $P=0.8$ (blue circles and lines)
and $P=0.5$ (magenta circles and lines). The pure optical-transport
calculations were then smeared using simple parametrisations of electronic
noise and cross talk in the MAPMT. The resultant distribution at $P=0.8$
is displayed with blue squares and dashed lines. 

The mean amplitudes in plots a) and c), which are the mean channels
of the pulse-height spectra, have been normalised so that the highest
mean amplitude in each distribution is 1.0. The relative widths plotted
in b) and d) are the ratios of the standard deviations and means of
the pulse-height spectra.

The simple model of electronic cross talk assumes that signal leakage
from one pixel to a nearest neighbour can be approximated by sampling
from a Gaussian distribution centred at 3\% of the signal amplitude
(number of scintillation photons) with a width ($\sigma)$ of 3\%
of the signal amplitude. Measured cross-talk values for the H12700
\citep{cross-talk} range from $\sim1\%$ to $\sim7\%$. For each
nearest neighbour, a randomly-generated leakage signal is then subtracted
from the pixel signal and added to the neighbour. For pixel $P_{j}$,
the nearest neighbours are $P_{j-8},P_{j-1},P_{j+1},P_{j+8}$ (if
the neighbour exists). Actual cross talk \citep{cross-talk} is rather
more complicated as the magnitude varies considerably from pixel to
pixel, the spectrum is not Gaussian and the cause can be both misrouting
of photoelectrons through the dynode chain and pickup effects at the
electronic output side. None the less the model gives an idea of the
magnitude of the effect.

Smearing of the signal due to noise has been sampled from a Gaussian
of width ($\sigma$) obtained from the measured widths of the pedestals
of the ADC spectra \citep{VME-daq}. The widths are equivalent to
$\sim2$ detected photons and at this level the effect of noise is
small compared to cross talk.

Comparing measurements with the base-level simulation, the latter
underpredicts the spreading of scintillation light from its point
of origin. It gives a position-dependent distribution of mean pulse
height which is too flat close to a pixel centre, and too steep close
to a pixel boundary. When imperfect polish is introduced, this increases
the scintillation-photon spreading. A value $P=0.5$ overpredicts
the spreading to adjacent pixels while $P=0.8$ is closer to the measurement,
although it still gives a distribution which is too flat near the
pixel centre. When 3\% cross talk (see above) is introduced at $P=0.8$
the apparent signal spreading is increased significantly and is similar
to the $P=0.5$ case with no cross talk. Thus it is difficult to separate
the two effects through comparison with the present measurements. 

Non-uniform sensitivity of the photocathode, as investigated for the
similar H8500 MAPMT \citep{mapmt-uniform}, is a further potential
source of distortion of the measured response. It has not been included
in the present calculations, pending a better knowledge of the sensitivity
of the employed H12700A MAPMT.

\subsection{\label{subsec:backgroundSim}Simulation of Potential Background}

Fast neutrons and gamma rays are a potential source of background
at the ESS. Here the measured and simulated response of a SoNDe module
to these backgrounds is presented. The measurements used the same
experimental setup as for the proton scan (Sec. \ref{subsec:Monoenergetic Protons})
and were made on the same day using the facilities of the Lund University
Source Test Facility \citep{STF}.

\subsubsection{\label{subsec:n-g-sources}Neutron and Gamma-ray Source Measurements}

\begin{figure}[h]
\begin{center}\includegraphics[width=0.6\columnwidth]{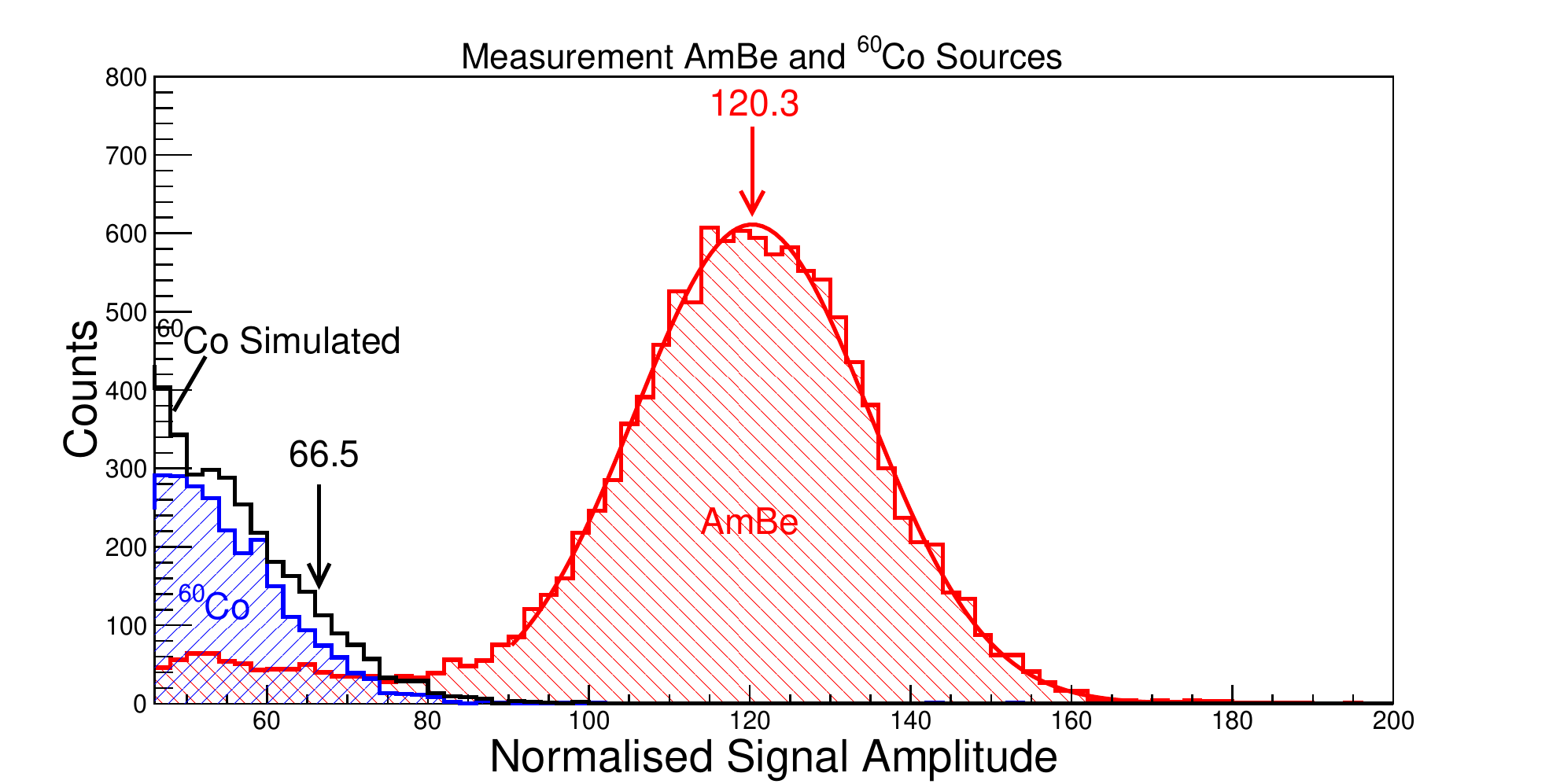}\end{center}

\caption{\label{fig:Pulse-Height-n-g}Pulse Height Measurement. Blue: $\mathrm{^{60}Co}$
gamma-ray source, red: AmBe neutron and gamma-ray source. The smooth
curve is a Gaussian fit to the measured neutron-capture peak. The
black line shows the end-point of the simulated $\mathrm{^{60}Co}$
spectrum.}
\end{figure}

The SoNDe module was irradiated with AmBe neutron and $^{60}\mathrm{Co}$
gamma-ray sources. A Pb brick of thickness 20~mm, placed between
the AmBe source and detector, attenuated gamma rays. A $\mathrm{CH_{2}}$
block of thickness 50~mm, placed detector-side of the Pb, provided
some moderation of the fast neutrons. The signal amplitude (Fig.~\ref{fig:Pulse-Height-n-g})
has been reconstructed from a nine-pixel cluster sum centred on pixel
P37. The AmBe run shows a prominant thermal-capture peak (energy 4.78~MeV)
at channel 120.3, which has been normalised to coincide with the simulated
number of detected scintillation photons. Measurements of the SoNDe
response to a collimated thermal neutron-beam produced at a reactor
\citep{n_IFE} show that the capture peak is very similar to that
obtained with AmBe. Background in the AmBe spectrum from fast neutrons
and gamma rays does not pollute the neutron-capture peak significantly.
This peak may thus be compared to that of the thermal-neutron simulation
(Fig.~\ref{fig:Neutron-response}), which does not include the fore-mentioned
background.

The $^{60}\mathrm{Co}$ spectrum shows the end point of the smeared
Compton distribution and compares it to the simulated distribution
(Sec.~\ref{subsec:Thermal-and-Fast}). The simulated edge (mean energy
$\sim1.0$~MeV) from the unresolved 1.17 and 1.33~MeV gamma-rays
aligns with channel $66.5$ while the slope of the measured spectrum,
normalised on the same basis as the neutron-capture peak, falls systematically
$\sim4$ channels lower. Thus the edge position for the measured $^{60}\mathrm{Co}$
spectrum is estimated to fall at channel $63\pm6.$

Compared to Compton electrons, the alpha-triton final state obviously
produces less scintillation light per unit energy deposited, which
is modeled in the simulation via the Birks formalism (Eq.~\ref{eq:2_birks}).

\subsubsection{\label{subsec:Thermal-and-Fast}Simulated Neutron and Gamma-ray Response}

\begin{figure}[h]
\includegraphics[width=1\columnwidth]{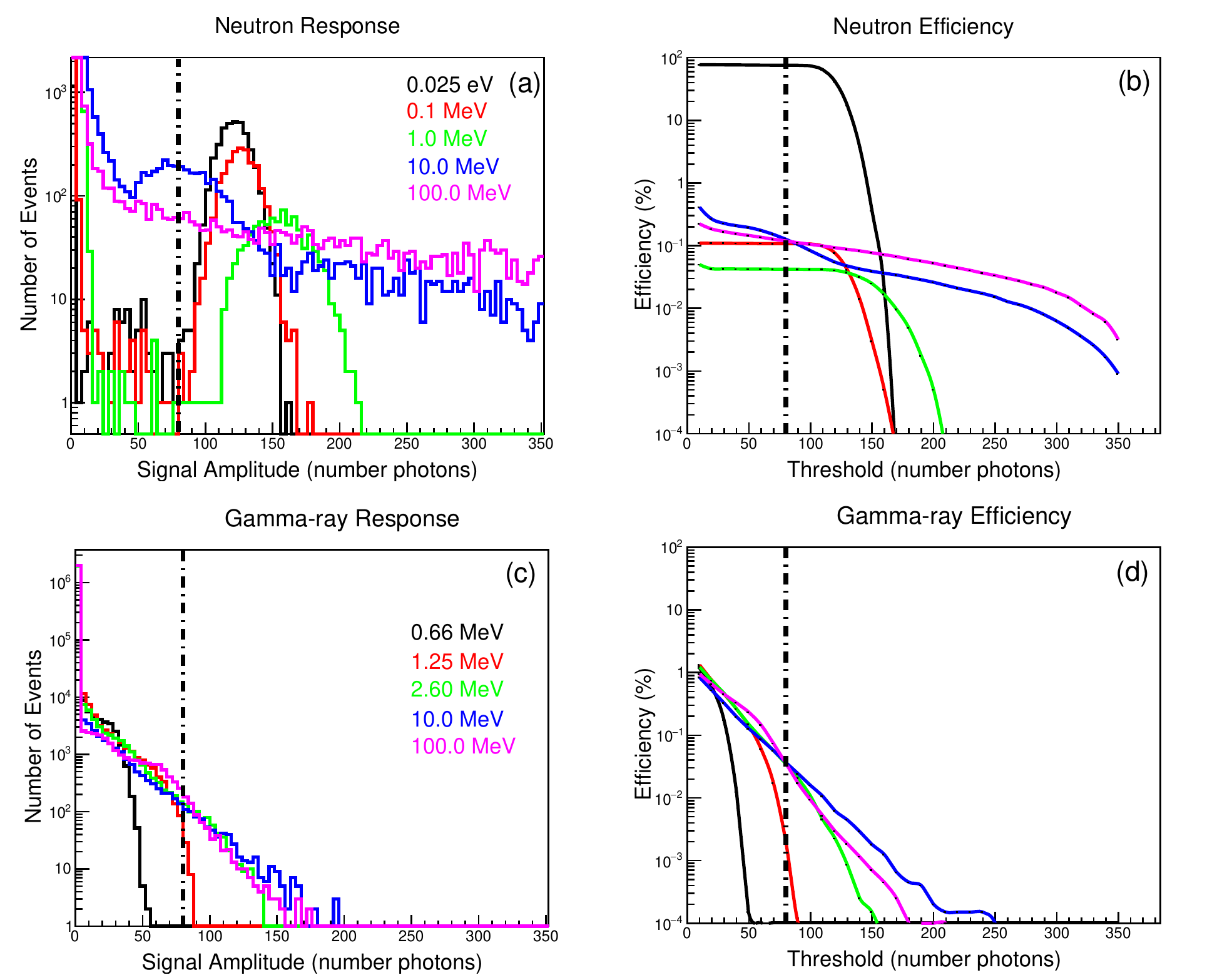}

\caption{\label{fig:Neutron-response}(a) Fast-neutron, signal amplitude compared
to that of 0.025~eV neutrons. (b) Neutron efficiency as a function
of threshold. The colour coding is as for plot (a). (c) Gamma-ray
signal amplitudes for a range of incident energies. (d) Gamma-ray
efficiency as a function of threshold where the colour code is as
plot (c). The dot-dash line in all panels marks a threshold level
of 80 photons. Note that the thermal-neutron spectrum resulted from
$5\times10^{3}$ incident particles, while the fast-neutron and gamma-ray
spectra resulted from $2\times10^{6}$ incident particles.}
\end{figure}
The simulated signal amplitude, in terms of number of scintillation
photons, for fast neutrons at incident energies from 0.1~MeV to 100~MeV
is compared to the thermal-neutron (0.025~eV) response in Fig.~\ref{fig:Neutron-response}(a).
The displayed pulse heights are cluster sums of the 9 pixel signal
amplitudes centred on P37. At low energies, where the fast neutron
has insufficient energy to fragment the nuclei of the glass, there
is a weak signal from neutron capture on $^{6}\mathrm{Li}.$ Where
the neutron has sufficient energy to break up nuclei, a continuum
emerges from the interactions of final state protons and light ions,
becoming more pronounced as the energy increases.

Neutron detection efficiency, defined as the ratio of integrated counts
above threshold to the total number of incident neutrons, is displayed
as a function of threshold in Fig.~\ref{fig:Neutron-response}(b).
The 80-photon threshold level results in insignificant loss of thermal
efficiency (76\%). At fast-neutron energies between 0.1, 1.0~MeV,
this threshold does not give significant background suppression, while
at higher energies it offers limited suppression. Neutron detection
efficiency at the 80-photon threshold is given in Table~\ref{tab:Detection-efficiency},
for a range of energies. A fast-neutron efficiency of around 0.1\%
is similar to that obtained with a 2.54~cm diameter $^{3}\mathrm{He}$
proportional counter \citep{he3-b10-fast-n,he3-fast-n} operating
at 10 bar, but around a factor $10^{2}$ larger than the $^{10}\mathrm{B}$-coated
`Multi Blade' detector \citep{fnSens} developed for ESS.

The signal-amplitude spectra of the SoNDe module at incident gamma-ray
energies from 0.66~MeV to 100~MeV have also been simulated. Fig.~\ref{fig:Neutron-response}(c)
displays the spectra, in terms of number of scintillation photons,
where the signal amplitude was reconstructed from a nine-pixel cluster
sum, as for neutrons. Gamma-ray spectra show a quite diffuse Compton
edge up to energies $\sim1$~MeV, but at higher energies secondary
electrons from Compton or pair-production processes have too much
energy to stop in $\sim1$~mm of glass and the distributions appear
quite similar from 2.6 up to 100~MeV. As the energy increases the
secondary electrons become more focused kinematically along the incident
gamma ray direction (perpendicular to the face of the GS20). This
tends to reduce the thickness of GS20 traversed and hence the signal
amplitude, so that the 10~MeV gamma-ray spectrum is enhanced at amplitudes
in the range 120 - 200 photons, compared to the 100~MeV gamma-ray
spectrum (Fig.\ref{fig:Neutron-response}(c)).

Gamma-ray detection efficiencies, as a function of detection threshold,
are given in Fig.~\ref{fig:Neutron-response}(d). A threshold of
80 photons (Table~\ref{tab:Detection-efficiency}) gives complete
suppression of low energy ($\leq1$ MeV) gamma rays and very effective
suppression at higher energies. The efficiency of 0.002\% for $^{60}\mathrm{Co}$
gamma rays (1.25~MeV average energy) appears higher than that measured
for a 2.54~cm, 10~bar $^{3}\mathrm{He}$ tube \citep{he3-fast-n},
but here the efficiency values are extremely sensitive to applied
thresholds.

\begin{table}[h]
\begin{center}%
\begin{tabular}{|c|c||c|c|}
\hline 
Gamma Source & Eff. (\%) & Neutron Source & Eff. (\%)\tabularnewline
\hline 
\hline 
0.66 MeV & 0.0 & 0.025 eV & 75.8\tabularnewline
\hline 
1.25 MeV & 0.002 & 0.1 MeV & 0.108\tabularnewline
\hline 
2.60 MeV & 0.035 & 1.0 MeV & 0.042\tabularnewline
\hline 
10.0 MeV & 0.037 & 10.0 MeV & 0.127\tabularnewline
\hline 
100.0 MeV & 0.036 & 100.0 MeV & 0.119\tabularnewline
\hline 
\end{tabular}\end{center}

\caption{\label{tab:Detection-efficiency}Detection efficiency for gamma rays
and neutrons at a threshold of 80 scintillation photons.}
\end{table}

The relative signal amplitude from thermal neutrons and gamma rays
has been calculated using a Birks factor $k_{B}$ (Eq.~\ref{eq:2_birks})
of 0.01 $\mathrm{mm/MeV}$. The neutron capture peak occurs at 120.3
photons. For $^{60}\mathrm{Co},$ the two gamma-ray lines have Compton
edges at 0.96 and 1.12 MeV, producing mean signal amplitudes of 61
and 72 photons respectively, with the mean of the two at 66.5 photons.
The simulated ratio of Compton edge to n-capture peak (0.55) is consistent
with the measured ratio of $0.52\pm0.05$ (Sec.~\ref{subsec:n-g-sources}). 

\subsubsection{\label{subsec:Comparison-with-proton}Comparison with proton data}

The measured neutron and gamma-ray pulse-height spectra are directly
comparable with the proton data, which were collected on the same
system, very close in time. For the vertical scan point closest to
the centre of P37, the measured 9-element sum shows a peak at channel
101.8, while the equivalent simulated value is 99.1. 

Taking the ratio of proton-peak pulse height to neutron-capture-peak
pulse height, the measured value is $0.85\pm0.02$, while the simulation
gives values of 0.82 at $k_{B}=0.01\:\mathrm{mm/MeV}$ and 0.93 at
$k_{B}=0.02\:\mathrm{mm/MeV}.$ 

Note that deuteron-beam scans are also reported in Ref.~\citep{Rofors_proton}
and there the proton and deuteron data are directly comparable. The
measured ratio of deuteron-to-proton pulse height \citep{Rofors_proton}
was 0.80 while the simulation gives 0.79 with $k_{B}=0.01\:\mathrm{mm/MeV}$
and 0.74 with $k_{B}=0.02\:\mathrm{mm/MeV}$. Thus the simulation
of proton, neutron-capture, deuteron and gamma-ray pulse height is
consistent with a value close to $0.01\:\mathrm{mm/MeV}$. Note that
direct comparison with the alpha-scan data was not possible as this
was performed with a different SoNDe module and DAQ system.

\section{\label{sec:Summary-and-Conclusion}Summary and Conclusions}

A Geant4-based simulation of the SoNDe module, designed for thermal-neutron
detection at ESS, is described. Transport of scintillation photons
through the glass scintillator and MAPMT window has been used to study
the spreading of the optical signal away from the interaction point
in the scintillator. This was performed for a variety of scintillator
grooving and optical coupling options. The likely experimental configuration,
with an ungrooved scintillator and no optical-coupling between the
scintillator and MAPMT, was found to give low collection efficiency
for the scintillation photons, but results in tighter containment
of the detected photons close to their point of origin, a goal of
the SoNDe design.

Simulations of the scintillation signal for different incident particles
have been compared to equivalent measurements made with alpha-particle,
gamma-ray and neutron sources, as well as beams of 2.5~MeV protons
and deuterons. 

An x-y scan, in 1~mm steps, of a collimated alpha-particle source
across the face of a SoNDe module was simulated and compared with
measurements. After collimation the illuminated `spot' on the GS20
was around 1.3~mm in diameter. Excellent agreement was obtained,
using the base-level simulation, on the drop in signal amplitude as
the interaction position moves away from the centre of a pixel of
the MAPMT. However the simulation tended to underestimate the degree
of light spreading slightly.

Finer-grained position scanning, in 0.5~mm steps, was implemented
using the LIBAF proton beam at Lund, which has a beam spot of $\sim100\:\mu$m
diameter on target. The measured position dependence of the scintillation
signal shows some discrepancy with the equivalent calculation of the
fall of the scintillation signal, as the interaction position is moved
away from a pixel centre. The base-level simulation, which assumes
perfectly polished reflecting surfaces, gives too small a variation
close to the pixel centre and too rapid a fall at the pixel border.
It underpredicts the amount of scintillation light spreading to adjacent
pixels. 

The simulation was extended to include non-perfect reflecting surfaces.
As the smearing from perfect specular reflection is increased, the
spreading of the optical photons from point of production also increases.
Similarly, the apparent amount of signal spreading increases when
the effects of electronic cross talk and noise in the MAPMT and associated
electronics are folded into the simulation. Much better correspondence
between simulation and measurement has been achieved using a surface
polish factor $P=0.8$ and assuming that on average 3\% of a pixel
signal crosses to its neighbour. However a more detailed knowledge
of the SoNDe module is required to quantify the relative contributions
of these effects. A measurement of cross talk and photocathode sensitivity,
for the actual MAPMT used in the measurements could be performed with
a fine-beam laser source. Although the GS20 surface appears smooth
to the naked eye, it would be desirable to investigate its structure
more microscopically, or alternatively to make detailed studies of
the reflection of a fine optical beam from points on the surface.

The signal from AmBe neutron and $\mathrm{^{60}Co}$ gamma-ray sources
was simulated and compared to measurements with the SoNDe module.
The relative strength of the signals from neutron capture and Compton
scattering of gamma rays are consistent with a Birks parameter of
0.01 $\mathrm{mm/MeV}$. Based on this comparison, the detection efficiency,
as a function of applied pulse-height threshold, has been simulated
for thermal neutrons, fast neutrons and gamma rays. A threshold equivalent
to 80 detected scintillation photons gives a thermal-neutron efficiency
of 75.8\%. Gamma rays up to 1~MeV are not detected at this threshold
and at higher energies have low efficiencies ($<0.04\%$). The fast-neutron
efficiency is somewhat higher at the 0.1\% level, which is comparable
to a typical $^{3}\mathrm{He}$ tube but a factor $\sim10^{2}$ larger
than a $^{10}\mathrm{B}$ -coated thermal-neutron detector.

A comparison between simulation and measurement of the relative pulse
height from neutron-capture, $\sim1.0$~MeV gamma rays, 2.5~MeV
protons and 2.5~MeV deuterons is consistent with a Birks parameter
of $k_{B}=0.01\mathrm{\:mm/MeV}$. However the general validity of
this value for a broader range of particle types and differential
energy loss $dE/dx$ remains to be investigated.

\section*{Acknowledgements}

We thank the following organisations for supporting this work: the
UK Science and Technology Facilities Council (Grant No. ST/P004458/1),
the UK Engineering and Physical Sciences Research Council Centre for
Doctoral Training in Intelligent Sensing and Measurement (Grant No.
EP/L016753/1), the European Union via the Horizon 2020 Solid-State
Neutron Detector Project (Proposal ID 654124), and the BrightnESS
Project (Proposal ID 676548). We also thank the LIBAF team at Lund
University for the provision and operation of their excellent proton
beam. The provision of detector and electronics hardware by the SoNDe
collaboration and the Lund Source Test Facility is gratefully acknowledged.


\begin{thebibliography}{99}
\bibitem{ess1}R. Garoby \emph{et al.}, Physica Scripta, 93(1) (2017),
014001.

\bibitem{ess2}S. Peggs \emph{et al}., ESS Technical Design Report
{[}ESS-2013-0001{]}. Technical report, European Spallation Source,
04 2013.

\bibitem{ess3}K. H. Andersen \emph{et al.}, Nucl. Instr. and Meth.
A 957 (2020), 163402.

\bibitem{skadi}S. Jaksch \emph{et al}., Nucl. Instr. and Meth. A
762 (2014), 22. \url{https://arxiv.org/abs/1403.2534}

\bibitem{SoNDe}S. Jaksch \emph{et al.}, arXiv 1707.08679, 2017. \url{https://arxiv.org/pdf/1707.08679}.

\bibitem{SoNDe2}S. Jaksch \emph{et al.}, \textquotedbl Recent Developments
of SoNDe High-Flux Detector Project\textquotedbl , Proc. Int. Conf.
on Neutron Optics (NOP2017), JPS Conf. Proc. 22 (2018), 011019.

\bibitem{SoNDe3}`Scintillator detector with high count rate', EPO
patent application 102014224449.8. 

\bibitem{GS20}GS20 scintillating glass, \url{https://scintacor.com/products/6-lithium-glass/}.

\bibitem{MAPMT}Hamamatsu H12700 Series Flat Panel Multianode PMT
assembly, \url{https://www.hamamatsu.com/resources/pdf/etd/H12700A_TPMH1348E.pdf}

\bibitem{ideas}IDEAS Integrated Detector Electronics AS, \url{https://ideas.no}.

\bibitem{Rofors_proton}E. Rofors \emph{et al.}, Nucl. Instr. and
Meth A 984 (2020), 164604.

\bibitem{Rofors-Alpha}E. Rofors \emph{et al}., Nucl. Instr. and Meth.
A 929 (2019), 90.

\bibitem{n_IFE}E. Rofors \emph{et al.}, arXiv:2010.0637 {[}physics.ins-det{]}
(2020).

\bibitem{G4}J. Allison \emph{et al}., Nucl. Instr. and Meth. A 835
(2016) 186-225, and references therein

\bibitem{G4Phys}Geant4 Physics Reference Manual, Release 10.6, 2020,
\url{http://geant4-userdoc.web.cern.ch/geant4-userdoc/UsersGuides/PhysicsReferenceManual/fo/PhysicsReferenceManual.pdf}

\bibitem{G4main}Geant4 Book for Application Developers, Release 10.6,
2020\url{http://geant4-userdoc.web.cern.ch/geant4-userdoc/UsersGuides/ForToolkitDeveloper/fo/BookForToolkitDevelopers.pdf}

\bibitem{Polish-fact}P. Gumplinger, Optical Photon Processes in Geant4,
Users' Workshop at CERN, Nov. 2002.

\bibitem{EJ-500}Optical Cement EJ-500, \url{https://eljentechnology.com/products/accessories/ej-500}.

\bibitem{Birks}J. B. Birks, The Theory and Practice of Scintillation
Counting, Pergamon Press, Oxford U.K., 1964.

\bibitem{EJ-520}$\mathrm{TiO_{2}}$ based reflective paint, \url{https://eljentechnology.com/products/accessories/ej-510-ej-520}.

\bibitem{Al-refl}G. Hass and J. E. Waylonis, J. Opt. Soc. Am. 51
(1961), 719.

\bibitem{GS20-1}M.P. Hehlen \emph{et al}., J. Appl. Phys. 124 (2018),124502.

\bibitem{borosilicate}Optical Glass Data Sheets, Schott AG Advanced
Optics, \url{www.schott.com/advanced_optics }.

\bibitem{ROOT}R. Brun and F. Rademakers, Nucl. Instr. and Meth. A
389 (1997) 81. \url{http://root.cern.ch/}.

\bibitem{gamSens}A. Khaplanov \emph{et al.}, J. Inst. 8 (2013), 10025,
\url{https://arxiv.org/abs/1306.6247}

\bibitem{fnSens}G. Mauri \emph{et al.}, J. Inst. 13 (2018), 03004,
\url{https://arxiv.org/abs/1712.05614 }

\bibitem{Tretyak}V. I. Tretyak, Astroparticle Phys. 33 (2010), 40.

\bibitem{LIBAF}Lund Ion Beam Analysis Facility (LIBAF), \url{http://www.nuclear.lu.se/forskning/tillaempad-kaernfysik/libaf/ }.

\bibitem{VME-daq}J. Scherzinger \emph{et al}., Appl. Radiation and
Isotopes 127 (2017), 98.

\bibitem{cross-talk}M. Calvi \emph{et al.}, J. Inst. 10 (2015), 09021,
\url{https://arxiv.org/abs/1506.04302}

\bibitem{mapmt-uniform}R. A. Montgomery \emph{et al}., Nucl. Instr.
and Meth. A 695 (2012), 326.

\bibitem{STF}F. Messi \emph{et al.}, \textquotedbl The neutron tagging
facility at Lund University\textquotedbl , IAEA Technical Report
on Modern Neutron Detection (2017), \url{https://arxiv.org/abs/1711.10286}

\bibitem{he3-b10-fast-n}G. Mauri \emph{et al.}, Eur. Phys. J. Tech.
and Instr. 6 (2019), Article 3, \url{ https://arxiv.org/abs/1902.09870v1}

\bibitem{he3-fast-n}F. Piscitelli \emph{et al}., Eur. Phys. J. Plus
135 (2020), 577, \url{https://arxiv.org/abs/2002.08153v1}
\end{thebibliography}
\end{document}